\documentclass[prc,aps,amsfonts,a4paper,nofootinbib,showpacs,twocolumn,floatfix]{revtex4}
\usepackage{graphicx}
\usepackage{bm}
\usepackage{hyperref}
\usepackage{amssymb}
\usepackage{amsfonts}
\usepackage{amsmath}
\usepackage{epstopdf}
\usepackage{natbib}
\usepackage{color}

\newcommand{\Eq}[1]{Eq.~(\ref{#1})}
\newcommand{\Eqs}[1]{Eqs.~(\ref{#1})}
\newcommand{\Fig}[1]{Fig.~\ref{#1}}
\newcommand{\Ref}[1]{Ref.~\cite{#1}}
\newcommand{\Sec}[1]{Sec.~\ref{#1}}
\newcommand{\eff}{\text{eff}}
\newcommand{\liq}{\text{liq}}
\newcommand{\gas}{\text{gas}}
\newcommand{\Sk}{\text{Sk}}
\newcommand{\ex}{\text{ex}}
\newcommand{\WS}{\text{WS}}
\DeclareMathOperator{\lap}{\nabla^2}
\DeclareMathOperator{\grad}{\bm{\nabla}}

\begin{document}
\title{Liquid-gas coexistence vs. energy minimization with respect to
  the density profile in the inhomogeneous inner crust of neutron stars}
\author{No\"el Martin} \email{noelmartin@ipno.in2p3.fr}
\author{Michael Urban} \email{urban@ipno.in2p3.fr}
\affiliation{Institut de Physique Nucl\'eaire, CNRS/IN2P3 and
  Universit\'e Paris-Sud, F-91406 Orsay, France} 
  \pacs{26.60.-c}

\begin{abstract}
  We compare two approaches to describe the inner crust of neutron
  stars: on the one hand, the simple coexistence of a liquid
  (clusters) and a gas phase, and on the other hand, the energy
  minimization with respect to the density profile, including Coulomb
  and surface effects. We find that the phase-coexistence model gives
  a reasonable description of the densities in the clusters and in the
  gas, but the precision is not high enough to obtain the correct
  proton fraction at low baryon densities. We also discuss the surface
  tension and neutron skin obtained within the energy minimization.
\end{abstract}

\maketitle

%%%%%%%%%%%%%%%%%%%%%%%%%%%%%%%%%%%%%%%%%%%%%%%%%%%%%%%%%%%%%%%%%%%%%%%%%%%%%%
\section{Introduction}
\label{sec:intro}
%%%%%%%%%%%%%%%%%%%%%%%%%%%%%%%%%%%%%%%%%%%%%%%%%%%%%%%%%%%%%%%%%%%%%%%%%%%%%%
The usual picture of the neutron-star crust \cite{Chamel2008} is that
one starts from a Coulomb crystal of nuclei in the outer crust. As one
goes deeper into the star, the nuclei become more and more neutron
rich (as a consequence of the increasing electron chemical potential
and $\beta$ equilibrium) until the neutron drip line is reached. This
defines the transition to the inner crust, where the nuclei
(``clusters'') are embedded in a dilute gas of unbound
neutrons. Descending further into the star, one expects to find the
so-called ``pasta phases''
\cite{Ravenhall1983,Hashimoto1984,Lassaut1987,Oyamatsu1993,Watanabe2003,Napolitani2007,Avancini2008,Avancini2010,Sebille2011,Baldo2014,Pais2014,Pais2015},
before one finally reaches the neutron-star core where matter
becomes homogeneous. The inhomogeneous phases in the inner crust can also be
interpreted in another way \cite{Avancini2008}, which is quite common in the
study of supernova matter (i.e., matter at finite temperature and out of $\beta$
equilibrium) \cite{Pais2015,Aymard2014}, namely as a consequence of
the first-order liquid-gas instability of nuclear matter. One aim of the present
paper is to see to what extent the simple picture of phase coexistence can
explain certain properties of the inhomogeneous phases of the inner crust of
neutron stars.

In particular, the phase coexistence picture corresponds to a
hydrostatic equilibrium which can serve as a starting point for a
hydrodynamic model of the collective modes in the inner crust
\cite{Magierski2004,MagierskiBulgac2004,DiGallo2011}. The collective
modes can have some impact on thermodynamic and transport properties
of the crust and may have observable consequences on the cooling of
the star \cite{PageReddy2012}. While there exist several calculations
of the collective modes in the crystalline phases
\cite{Khan2005,Cirigliano2011,Chamel2013CM}, it seems to be difficult
to model the collective modes in the pasta phases beyond the
hydrodynamic approach \cite{DiGallo2011}.

Within the most simple phase-coexistence picture, surface- and Coulomb
energies are neglected. An approximate way to include them is the
compressible liquid-drop model (see, e.g., \cite{Pais2015}). However,
this model does not account for the neutron skin and for the effect of
the surface diffuseness on the Coulomb energy. Therefore, we will use
here a more microscopic approach, namely to parameterize the density
profile and determine its parameters by minimizing the thermodynamic
potential. Nevertheless it will turn out that the densities in the
dense (``liquid'') and dilute (``gas'') regions satisfy quite well the
conditions of phase coexistence.

The article is organized as follows. In Sec.~\ref{sec:formalism}, we
describe the formalisms for the phase coexistence and for the more
complete minimization of the thermodynamic potential. In
Sec.~\ref{sec:results}, we show and compare the results obtained in
both models, and in Sec.~\ref{sec:conclusions} we conclude.
 
%%%%%%%%%%%%%%%%%%%%%%%%%%%%%%%%%%%%%%%%%%%%%%%%%%%%%%%%%%%%%%%%%%%%%%%%%%%%%%
\section{Formalism}
\label{sec:formalism}
%%%%%%%%%%%%%%%%%%%%%%%%%%%%%%%%%%%%%%%%%%%%%%%%%%%%%%%%%%%%%%%%%%%%%%%%%%%%%%
\subsection{Phase coexistence}
%%%%%%%%%%%%%%%%%%%%%%%%%%%%%%%%%%%%%%%%%%%%%%%%%%%%%%%%%%%%%%%%%%%%%%%%%%%%%%
As a first approximation, the inner crust could be described as a phase
coexistence of liquid drops (corresponding to the nuclear clusters) with volume
$V^{\liq}$ and a gas (the dilute neutron gas) with volume $V^{\gas}$, which
satisfies mechanical and chemical equilibrium, i.e.,
\begin{align}
  P^{\gas} &= P^{\liq} \,, \label{eq:press_equil}\\
  \mu^{\gas}_q &= \mu^{\liq}_q \,, \label{eq:chem_equil}
\end{align}
with $P^i$ the pressure, and $\mu_q^i$ the chemical potential of
neutrons ($q=n$) or protons ($q=p$), respectively, in the phase $i$.

In addition to the neutrons and protons, we consider a uniform
electron gas to ensure charge neutrality, i.e., $V \rho_e = V^{\liq}
\rho^{\liq}_p + V^{\gas} \rho^{\gas}_p$, where $V = V^{\liq}+V^{\gas}$
is the total volume and $\rho_p^i$ is the proton density in phase $i$
($\rho_p^{\gas}$ is not always zero). Instead of working with the
volumes, it is more convenient to introduce the volume fraction $u$
filled by the liquid, which satisfies
\begin{equation}
  u = \frac{V^{\liq}}{V} 
    = \frac{\rho_e-\rho_p^{\gas}}{\rho_p^{\liq}-\rho_p^{\gas}} \, .
  \label{eq:volume_ratio}
\end{equation}

Furthermore, in neutron stars, matter is in $\beta$ equilibrium, i.e.,
\begin{equation}
  \mu_e = \mu_n - \mu_p \, \label{eq:beta_equil}
\end{equation}
with $\mu_e$ the chemical potential of the electrons. For practical
purposes, the electron mass can be neglected so that $\mu_e =
(3\pi^2\rho_e)^{1/3} \hbar c$. We thus obtain for the volume fraction
\begin{equation}
  u = \frac{1}{\rho_p^{\liq} - \rho_p^{\gas}} \left[
    \frac{(\mu_n-\mu_p)^3}{3 \pi^2(\hbar c)^3} - \rho_p^{\gas}
    \right] \, .
    \label{eq:volume_ratio_full}
  \end{equation}

We use a Skyrme energy-density functional (EDF) \cite{Chabanat1997} to
calculate the energy density $\epsilon_{\Sk}$ as a function of $\rho_n$ and
$\rho_p$. This allows us to write explicit expressions for the
chemical potentials $\mu_q = \partial \epsilon_{\Sk}/\partial\rho_q$
and the pressure $P = -\epsilon_{\Sk} + \mu_n\rho_n+\mu_p\rho_p$
appearing in \Eqs{eq:chem_equil} and (\ref{eq:press_equil}). We use
different Skyrme parametrizations, all fitted to the neutron-matter
equation of state: the Saclay-Lyon forces SLy4 and SLy7
\cite{Chabanat1998} and the Brussels-Montreal forces BSk20 and BSk22
\cite{Chamel2013}.

%%%%%%%%%%%%%%%%%%%%%%%%%%%%%%%%%%%%%%%%%%%%%%%%%%%%%%%%%%%%%%%%%%%%%%%%%%%%%%
\subsection{Energy minimization}
\label{sec:formalism_etf}
%%%%%%%%%%%%%%%%%%%%%%%%%%%%%%%%%%%%%%%%%%%%%%%%%%%%%%%%%%%%%%%%%%%%%%%%%%%%%%
In the phase-coexistence picture described above, we could determine
the volume fraction $u$, but not the actual size of the clusters. The
latter is determined by finding the best compromise between the
Coulomb energy (favoring small clusters) and the surface energy
(favoring large clusters), which were both neglected in the
phase-coexistence picture. Let us now consider a more sophisticated
approach, where these effects are included.

The surface energy is computed within the semiclassical Extended
Thomas-Fermi (ETF) approximation \cite{Brack1985,Aymard2014}. Instead
of solving Euler-Lagrange equation \cite{Lassaut1987,Baldo2014}, we
use a parametrization of the density profile and determine the
parameters by minimizing the energy. Concerning the Coulomb energy, we
use the Wigner-Seitz (WS) approximation, i.e., we calculate the
Coulomb energy in an isolated cell $r<R_{\WS}$, with $R_{\WS}$
chosen such that the volume of the WS cell corresponds to the volume
of the unit cell. Inside the WS cell, our density is parametrized as
\begin{equation}
  \rho_q(r) = \rho_q^{\gas} + \frac{\rho_q^{\liq}-\rho_q^{\gas}} 
    {1 + \exp \left[(r-r_q)/a_q \right]}
    \label{eq:density_profile}
\end{equation}
and the energy is minimized with respect to the 9 parameters
$\rho_q^i$, $r_q$, $a_q$, and $R_{\WS}$. This parametrization looks
similar to the shape obtained in Hartree-Fock calculation, e.g.,
\Ref{Negele1973} and has a simple interpretation: $\rho_q^{\liq}$ and
$\rho_q^{\gas}$ correspond, respectively, to the asymptotic densities
in the cluster and in the gas far away from the surface, $r_q$
describes the cluster radius, including the possibility of a neutron
skin if $r_n > r_p$, and $a_q$ is the surface diffuseness. We also
tried a more general parametrization using two more parameters,
namely, raising the Fermi function in \Eq{eq:density_profile} to a
power $\gamma_q$. This additional degree of freedom allows one to
describe an ``asymmetric'' surface, but since it did not substantially
change our results we discarded it for simplicity.

Note that a similar approach was followed by Oyamatsu
\cite{Oyamatsu1993} and by Pearson et al. \cite{Pearson2012}. However,
Oyamatsu used a completely different parametrization of the density,
which did not become constant inside the cluster. Our parametrization
(\ref{eq:density_profile}) resembles more the one used by Pearson et
al.

It seems very hard to perform the minimization of the energy for fixed
average densities (averaged over the WS cell), since the average
densities depend on all the parameters. We therefore introduce
chemical potentials $\mu_q$ as Lagrange parameters to fix the
densities and minimize the thermodynamic potential instead of the
energy. To be precise, we minimize $\omega = -P = \Omega / V_{\WS}$,
where
\begin{equation}
  \Omega = E_{\Sk} - \mu_n N - \mu_p Z - \mu_e Z + E_e + E_C + E_{\ex}\,,
    \label{eq:omega_expand}
\end{equation}
which reduces to
\begin{equation}
  \Omega = E_{\Sk} - \mu_n (N + Z) + E_e + E_C + E_{\ex}\,,
  \label{eq:omega}
\end{equation}
because of the $\beta$-equilibrium constraint (\ref{eq:beta_equil}).
In this equation, $E_{\Sk}$ denotes the energy obtained with the
Skyrme functional, $N$ and $Z$ are the total numbers of neutrons and
protons in the WS cell, $E_e$ denotes the energy of the electron gas,
$E_C$ the Coulomb energy, and $E_{\ex}$ the energy due to the Coulomb
exchange term.

The volume $V_{\WS}$ of the WS cell depends on $R_{\WS}$ and on the
geometry one considers (spheres, rods, or slabs). Let us define the
functions $S_d(r)$ and $V_d(r)$ describing, respectively, the
``surface'' and the ``volume'' of a $d$ dimensional sphere of radius
$r$, i.e.,
\begin{align}
S_1(r) &= 2\,,& S_2(r) &= 2\pi r\,,& S_3(r) &= 4\pi r^2\,, 
\label{eq:s_d}
\\
V_1(r) &= 2r\,,& V_2(r) &= \pi r^2\,,& V_3(r) &= \tfrac{4}{3}\pi r^3\,.
\label{eq:v_d}
\end{align}
In the case of spheres ($d=3$), the volume of the WS cell is $V_{\WS}
= V_3(R_{\WS})$. In the case of rods ($d=2$), the ``volume'' $V_{\WS}
= V_2(R_{\WS})$ is actually an area and consequently $E$, $N$, $Z$,
etc. represent energies and particle numbers per unit
length. Similarly, in the case of slabs ($d=1$), the ``volume''
$V_{\WS} = V_1(R_{\WS})$ is a length and $E$, $N$, $Z$, etc. represent
energies and particle numbers per unit area.

The integrated Skyrme energy in the WS cell, $E_{\Sk}$, is defined as
\begin{equation}
  E_{\Sk} = \int_0^{R_{\WS}} d^dr \, \mathcal{H}[\rho_n(r),\rho_p(r)] \,,
    \label{eq:enuc}
\end{equation}
with $d^dr = S_d(r) dr$ and $\mathcal{H}$ the Skyrme EDF detailed in
\Ref{Chabanat1997}, the kinetic energy density $\tau_q(r)$ being
calculated within the ETF approximation as described in
\cite{Brack1985,Aymard2014}. \footnote{Note that terms containing
  $\lap \rho_q$ pose numerical problems for $d=2$ and $3$ since the
  parametrization of the density (\ref{eq:density_profile}) is only
  approximately constant at the origin (in contrast to the density
  profile used in \Ref{Pearson2012}, whose derivatives vanish at $r=0$
  and $r=R_{\WS}$). We avoid these problems by integrating the terms
  of the functional containing $\lap\rho_q$ by parts and then using
  the form of the functional obtained in this way in the numerical
  calculation.}

For the energy of the electron gas, we assume a constant density
$\rho_e = Z/V_{\WS}$. Hence, the energy can be written as $E_e =
V_{\WS} \epsilon_e$. The energy density $\epsilon_e$, taking into
account the electron mass $m_e$, reads \cite{Baldo2014}
\begin{equation}
  \epsilon_e = \frac{\hbar c k_{F,e}^4}{4 \pi^2}
    \left[\left(1 + \frac{1}{2x_e^2}\right) \sqrt{1+\frac{1}{x_e^2}}
    - \frac{\sinh^{-1} x_e}{2x_e^4}\right] \,,
    \label{eq:electron_energy}
\end{equation}
with $k_{F,e} = (3\pi^2\rho_e)^{1/3}$ and $x_e = \hbar k_{F,e}/m_e c$.

The Coulomb energy $E_C$ is derived from the charge density $\rho_c(r)
= \rho_p(r)-\rho_e$. We first calculate the Coulomb potential $V_C(r)$
satisfying the Poisson equation $\lap V_C(r) = 4\pi e^2 \rho_c(r)$
with $\grad V_C(0) = 0$. This determines $V_C(r)$ only up to a constant,
which is however irrelevant for the energy since the total charge of
the WS cell is zero. The Coulomb energy is given by
\begin{equation}
  E_C = \frac{1}{2} \int_0^{R_{\WS}} d^d r \rho_c(r) V(r) \,.
    \label{eq:coulomb_energy}
\end{equation}
Note that $V_C(r)$ and the integral (\ref{eq:coulomb_energy}) must be
computed numerically, in contrast to the approximate expressions one
obtains for uniformly charged spheres (or rods, or slabs,
respectively) with a sharp surface which are often used in the
literature \cite{Ravenhall1983,Hashimoto1984,Avancini2010,Pais2014}.

We relaxed the assumption of a constant electron density by using a
screened Coulomb potential instead of the full one, i.e., replacing in
the Poisson equation $\lap V_C(r)$ by ($\lap - 1/\lambda^2) V_C(r)$,
where $\lambda$ denotes the Debye screening length, $1/\lambda^2 =
(4\alpha/\pi) k_{Fe}^2$ (neglecting the electron mass), with $\alpha =
e^2/\hbar c$. However, as already noticed in \cite{Maruyama2005}, the
screening length of the electrons is so large that this effect is
negligible and the calculations of the present paper were done without
electron screening.

The exchange term of the Coulomb interaction is computed within the
Slater approximation \cite{Baldo2014}. For the protons, it is given by
the integral of
\begin{equation}
  \epsilon_{\ex,p}(r) = - \frac{3}{4} \left(\frac{3}{\pi}\right)^{1/3}
    e^2 \rho_p^{4/3}(r) \,,
  \label{eq:exch_prot}
\end{equation}
over the WS cell. For the relativistic electrons, the exchange energy
includes also contributions from transverse photons and gets positive
for $x_e \gtrsim 2.53$ \cite{Salpeter1961}. Its expression reads
\cite{Rajagopal1978}
\begin{equation}
  \epsilon_{\ex,e} = \frac{e^2 k_{F,e}^4}{8 \pi^3}
    \left\{3\left[\sqrt{1 + \frac{1}{x_e^2}} - \frac{\sinh^{-1} x_e}{x_e^2} 
      \right]^2 - 2\right\}\,,
  \label{eq:exch_elec}
\end{equation}

%%%%%%%%%%%%%%%%%%%%%%%%%%%%%%%%%%%%%%%%%%%%%%%%%%%%%%%%%%%%%%%%%%%%%%%%%%%%%%
\section{Results}
\label{sec:results}
%%%%%%%%%%%%%%%%%%%%%%%%%%%%%%%%%%%%%%%%%%%%%%%%%%%%%%%%%%%%%%%%%%%%%%%%%%%%%%
\subsection{Phase coexistence}
\begin{figure}
  \includegraphics[width=8cm]{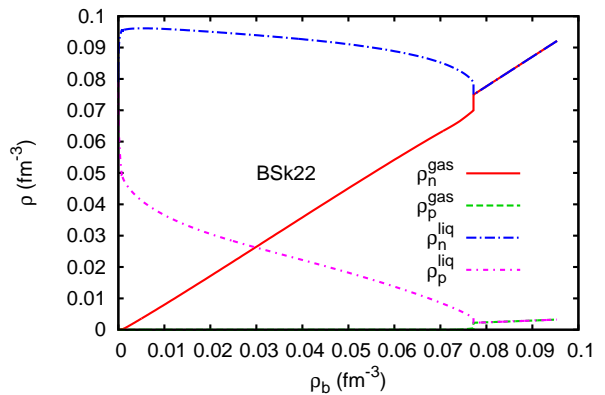}
  \caption{(Color online) Neutron and proton densities in the coexisting liquid
    (droplets) and gas phases, calculated with the Skyrme
    parametrization BSk22.}
  \label{fig:coex_density}
\end{figure}

%%%%%%%%%%%%%%%%%%%%%%%%%%%%%%%%%%%%%%%%%%%%%%%%%%%%%%%%%%%%%%%%%%%%%%%%%%%%%%
Let us first discuss the results obtained within the simple phase
coexistence approach. Figure~\ref{fig:coex_density}
%%%%%%%%%%%%%%%%%%%%%%%%%%%%%%%%%%%%%%%%%%%%%%%%%%%%%%%%%%%%%%%%%%%%%%%%%%%%%%
%%%%%%%%%%%%%%%%%%%%%%%%%%%%%%%%%%%%%%%%%%%%%%%%%%%%%%%%%%%%%%%%%%%%%%%%%%%%%%
represents the densities in the gas and in the liquid as functions of
the baryon density
\begin{equation}
  \rho_b = u(\rho_n^{\liq} + \rho_p^{\liq}) + (1-u) (\rho_n^{\gas} +
    \rho_p^{\gas}) \,,
    \label{eq:baryonic_density}
\end{equation}
obtained with the BSk22 interaction \cite{Chamel2013}. Results
calculated with other Skyrme parametrizations are close to those
displayed for BSk22 (cf.  Fig.~\ref{fig:miniSly4} for SLy4). The main
difference is the transition to uniform matter, i.e., the point where
the liquid fills the whole volume ($u = 1$). For instance, the
transition occurs at $\rho_b \approx 0.09~\text{fm}^{-3}$ with SLy4,
while with BSk22 it is below $0.08~\text{fm}^{-3}$.

\begin{figure}
  \includegraphics[width=8cm]{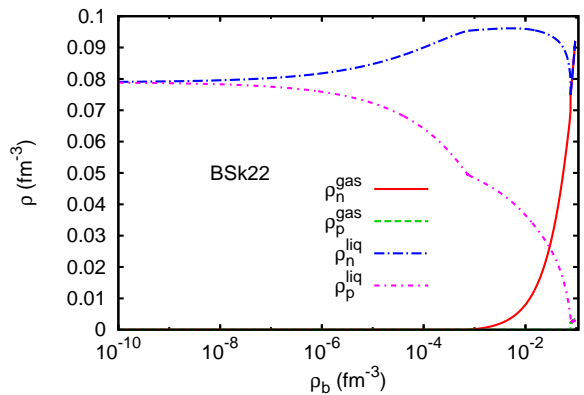}
  \caption{(Color online) Same as \Fig{fig:coex_density} but with a logarithmic scale
    for the baryon density.}
  \label{fig:coex_density_log}
\end{figure}

Figure~\ref{fig:coex_density_log}
%%%%%%%%%%%%%%%%%%%%%%%%%%%%%%%%%%%%%%%%%%%%%%%%%%%%%%%%%%%%%%%%%%%%%%%%%%%%%%
%%%%%%%%%%%%%%%%%%%%%%%%%%%%%%%%%%%%%%%%%%%%%%%%%%%%%%%%%%%%%%%%%%%%%%%%%%%%%%
shows the same results, but on a logarithmic scale so that the low
density region is better visible. The region below $\rho_b \approx
10^{-3}~\text{fm}^{-3}$ corresponds to the outer crust, where
$\rho^{\gas}_n = \rho^{\gas}_p=0$ (i.e., $\mu_n, \mu_p \leq 0$). In
the limit of extremely low densities, the droplets are made of
symmetric nuclear matter at saturation density ($\rho_n = \rho_p =
0.08$ fm$^{-3}$). The kink in the densities visible in
\Fig{fig:coex_density_log} corresponds to the transition between the
outer and the inner crust, i.e., to the point where $\mu_n$ becomes
positive and the neutron gas appears.

\begin{figure}
  \includegraphics[width=8cm]{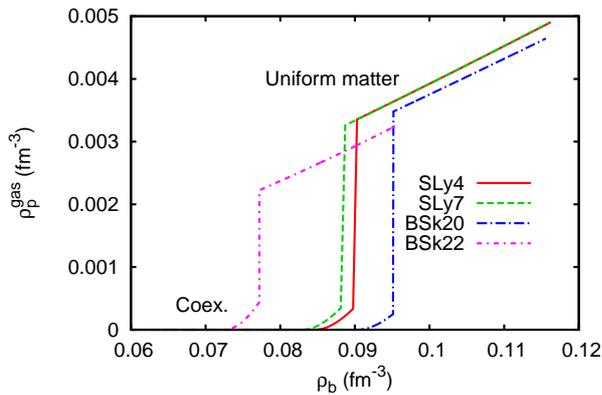}
  \caption{(Color online) Proton density in the gas phase (and later in uniform
    matter) as function of the total baryon density calculated within
    the phase coexistence approach using the SLy4, SLy7, BSk20 and
    BSk22 Skyrme parametrizations.}
  \label{fig:coex_density_prot}
\end{figure}

Close to the transition to uniform matter, also $\mu_p$ becomes
positive and the gas phase does not only contain neutrons, but also a
small amount of protons. Figure~\ref{fig:coex_density_prot}
%%%%%%%%%%%%%%%%%%%%%%%%%%%%%%%%%%%%%%%%%%%%%%%%%%%%%%%%%%%%%%%%%%%%%%%%%%%%%%
%%%%%%%%%%%%%%%%%%%%%%%%%%%%%%%%%%%%%%%%%%%%%%%%%%%%%%%%%%%%%%%%%%%%%%%%%%%%%%
zooms on densities where protons are present in the gas. Note that the
proton density in the gas is always very low, less than
$10^{-3}~\text{fm}^{-3}$ for all Skyrme parametrizations, compared to
the other densities calculated in the phase coexistence framework.

%%%%%%%%%%%%%%%%%%%%%%%%%%%%%%%%%%%%%%%%%%%%%%%%%%%%%%%%%%%%%%%%%%%%%%%%%%%%%%
\subsection{Energy minimization}
%%%%%%%%%%%%%%%%%%%%%%%%%%%%%%%%%%%%%%%%%%%%%%%%%%%%%%%%%%%%%%%%%%%%%%%%%%%%%%
In the preceding subsection, we treated the inner crust as two nuclear
fluids in phase coexistence. As already mentioned in
\Sec{sec:formalism_etf}, this approach misses surface and Coulomb
effects, which are included in the minimization of the thermodynamic
potential with respect to the parameters of the density profile given
in \Eq{eq:density_profile}. Actually, we perform the minimization for
each of the three different geometries discussed in
\Sec{sec:formalism_etf}, namely spheres (3D), rods (2D), and slabs
(1D). The geometry giving the lowest thermodynamic potential should be
the one that is physically realized.

\begin{figure}[htpb]
  \includegraphics[width=8cm]{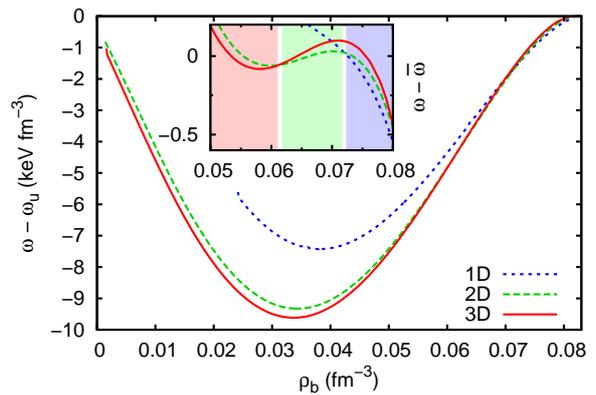}
  \caption{(Color online) The thermodynamic potential $\omega$ as function of the
    baryon density $\rho_b$, with the thermodynamic potential
    $\omega_u$ of uniform matter subtracted. In the inset, a
    phenomenological function $\overline{\omega}$ is subtracted, which
    approximates the average behavior of the three $\omega$'s.
    The SLy4 interaction was used in this calculation.}
  \label{fig:minim_energy}
\end{figure}

From now on, we restrict ourselves to the range $\mu_n > 0$
corresponding to the inner crust and show the results as functions of
the average baryon density $\rho_b = A/V_{\WS}$, where $A=N+Z$ is the
total number of nucleons in the WS cell. Figure~\ref{fig:minim_energy}
%%%%%%%%%%%%%%%%%%%%%%%%%%%%%%%%%%%%%%%%%%%%%%%%%%%%%%%%%%%%%%%%%%%%%%%%%%%%%%
%%%%%%%%%%%%%%%%%%%%%%%%%%%%%%%%%%%%%%%%%%%%%%%%%%%%%%%%%%%%%%%%%%%%%%%%%%%%%%
displays the difference $\omega-\omega_u$ between the thermodynamic potentials
$\omega$ in 3D, 2D, and 1D geometry, obtained by numerical minimization with the
SLy4 interaction, and the thermodynamic potential $\omega_u$ of uniform $npe$
matter in $\beta$ equilibrium with the same baryon density. The difference is
negative, confirming that the inhomogeneous phase is favored over uniform matter
in this density range. In order to make the differences between the 3D, 2D, and
1D geometries better visible, we subtract in the inset a purely phenomenological
function $\bar{\omega}$ which approximates the average behavior of the three
$\omega$'s. At densities below $0.061~\text{fm}^{-3}$, the most favorable phase
is the crystal (3D). From $0.061$ to $0.073~\text{fm}^{-3}$ the preferred phase
are the rods (``spaghetti'', 2D). Finally between $0.073$ and
$0.081~\text{fm}^{-3}$ we find the slabs (``lasagne'', 1D), until the system
transforms into uniform matter. In contrast to other work
\cite{Ravenhall1983,Hashimoto1984,Lassaut1987,Oyamatsu1993,Pais2014,Pais2015}, we did not
find ``inverted'' geometries such as tubes and bubbles (``Swiss cheese''), which
would correspond to $\rho_q^{\gas} > \rho_q^{\liq}$ in the parametrization
(\ref{eq:density_profile}) of the density profile in 2D and 3D, respectively.

Let us note that the energy differences between the three geometries
are extremely small compared to the total energy, especially between
2D and 3D beyond $\sim0.05~\text{fm}^{-3}$, so that one may expect
coexistence of different geometries. This is because, in contrast to
most studies of the pasta phases, we do not consider a fixed proton
fraction, but $\beta$ equilibrium. As pointed out in
\Ref{Piekarewicz2012}, the small proton fraction (see below)
corresponding to $\beta$ equilibrium is very unfavorable for the
formation of pasta phases. Our method results necessarily in
first-order phase transitions between the different geometries. In
reality, however, it might happen that the system passes continuously
from one phase to another, e.g., by deforming the nuclei in the 3D
phase before they merge into rods \cite{Ravenhall1983,Lattimer1991}.

\begin{figure}
  \includegraphics[width=8cm]{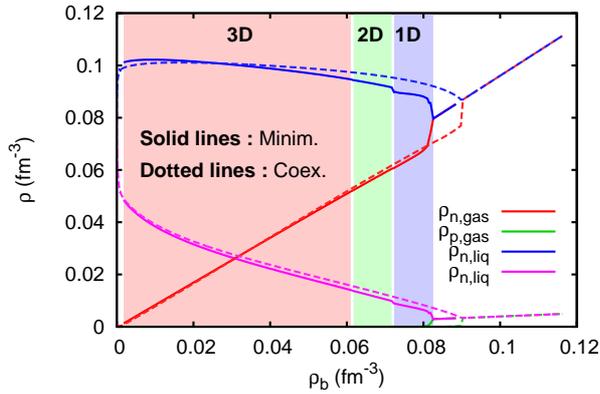}
  \caption{(Color online) Solid lines: densities in the cluster ($\rho_q^{\liq}$)
    and in the gas ($\rho_q^{\gas}$) corresponding to the geometry
    (3D, 2D, 1D, or uniform matter) that minimizes $\omega$ in
    \Fig{fig:minim_energy}, obtained by the minimization of the SLy4
    functional. Dots: results obtained within the phase
    coexistence approach as in \Fig{fig:coex_density} but with the
    SLy4 interaction.}
\label{fig:miniSly4}
\end{figure}

The densities $\rho_q^i$ in the favored geometry obtained by the
minimization are displayed in Fig.~\ref{fig:miniSly4} as the solid
lines.
%%%%%%%%%%%%%%%%%%%%%%%%%%%%%%%%%%%%%%%%%%%%%%%%%%%%%%%%%%%%%%%%%%%%%%%%%%%%%%
%%%%%%%%%%%%%%%%%%%%%%%%%%%%%%%%%%%%%%%%%%%%%%%%%%%%%%%%%%%%%%%%%%%%%%%%%%%%%%
We notice a discontinuity at the transition from 2D to 1D. It is due
to the first-order phase transition mentioned above (the corresponding
jump between 3D to 2D is too small to be seen on the figure). For
comparison, the dotted lines in Fig.~\ref{fig:miniSly4} were
calculated as in \Fig{fig:coex_density} within the phase coexistence
framework. The densities obtained within both approaches are quite
similar, which means that even with Coulomb and surface effects, the
mechanical and chemical equilibrium, \Eqs{eq:press_equil} and
(\ref{eq:chem_equil}), are approximately satisfied for the densities
in the cluster and in the gas away from the interface. The main
difference is the transition density from the inner crust to uniform
matter, i.e., to the neutron star core. Since Coulomb and surface
effects favor uniform matter, the transition happens earlier (i.e., at
lower $\rho_b$) in the minimization than in the phase coexistence
approach.

\begin{figure}
  \includegraphics[width=8cm]{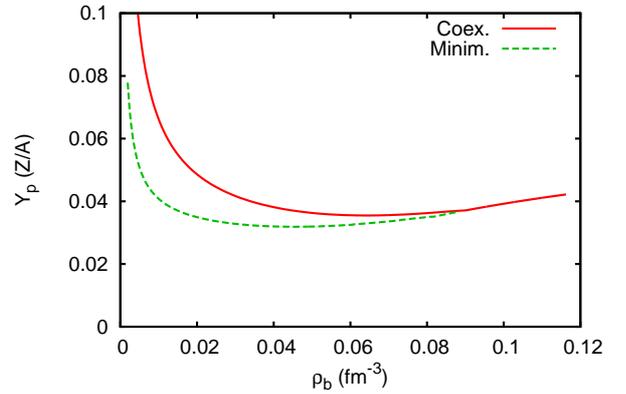}
  \caption{(Color online) Proton fraction $Y_p$ as function of the baryon density
    $\rho_b$. The proton fraction is computed using the SLy4
    interaction with the phase coexistence (solid line) and the energy
    minimization (dots) approach.}
  \label{fig:protfrac}
\end{figure}

Although the differences in the densities are small, they have a
sizable effect on the proton fraction $Y_p = Z/A$. This can be seen in
Fig.~\ref{fig:protfrac},
%%%%%%%%%%%%%%%%%%%%%%%%%%%%%%%%%%%%%%%%%%%%%%%%%%%%%%%%%%%%%%%%%%%%%%%%%%%%%%
%%%%%%%%%%%%%%%%%%%%%%%%%%%%%%%%%%%%%%%%%%%%%%%%%%%%%%%%%%%%%%%%%%%%%%%%%%%%%%
where we compare the proton fractions obtained within the phase coexistence
approach (solid lines) and by energy minimization (dots). It appears that at low
baryon density, the proton fraction obtained by minimizing the energy is
significantly lower than in the phase coexistence picture, although the proton
densities in the liquid are very close (cf.~\Fig{fig:miniSly4}). A very similar
difference between the two approaches is observed in the Relativistic Mean-Field
(RMF) framework \cite{Avancini2008}. This disagreement can be traced back to
the tiny difference in $\rho_n^{\gas}$ due to the small Coulomb and surface
corrections to the chemical and mechanical equilibrium. Since in this region the
total density is dominated by the density of the gas,
$\rho_b\approx\rho_n^{\gas}$, and the volume fraction $u$ is approximately
proportional to the difference $\rho_B-\rho_n^\gas$, the quantities $u$ and
consequently also $Y_p$ are very sensitive to small deviations of
$\rho_n^{\gas}$.

The proton fractions obtained by energy minimization are similar to
those obtained by Pearson et al. \cite{Pearson2014} within the
Hartree-Fock-Bogoliubov (HFB) model with BSk interactions. However, in
contrast to the HFB model, the ETF approximation does not include
shell effects. Therefore, our proton number in the 3D phase varies
smoothly from $Z\approx 34$ at low baryon density to $Z\approx 26$ at
the transition to the 2D phase, while the HFB model gives $Z=40$ in
the whole inner crust except in the case of the BSk22 interaction
where $Z$ jumps from 40 to 20 at $\rho_b = 0.035$ fm$^{-3}$
\cite{Pearson2014}.

%%%%%%%%%%%%%%%%%%%%%%%%%%%%%%%%%%%%%%%%%%%%%%%%%%%%%%%%%%%%%%%%%%%%%%%%%%%%%%
\subsection{Properties of the liquid-gas interface}
%%%%%%%%%%%%%%%%%%%%%%%%%%%%%%%%%%%%%%%%%%%%%%%%%%%%%%%%%%%%%%%%%%%%%%%%%%%%%%
Let us discuss in some more detail the properties of the liquid-gas
interface obtained by energy minimization. To that end, we have to
compare our WS cell with a reference system containing a cluster with
constant densities $\rho_q^{\liq}$ and a sharp surface, surrounded by
a gas with constant densities $\rho_q^{\gas}$. The presence of a
neutron skin $s_n = r_n-r_p > 0$ complicates this comparison, and we
follow Douchin and Haensel \cite{Douchin2000} who discussed this
problem in detail.

We define the radius $r_p^{\eff}$ of the reference cluster in such a
way that it contains the same number of protons as the actual WS cell,
i.e.,
\begin{equation}
(\rho_p^{\liq}-\rho_p^{\gas}) V_d(r_p^{\eff}) + \rho_p^{\gas} V_{\WS} = Z\,
\label{eq:defrpeff}
\end{equation}
($\rho_p^{\gas} = 0$ in most cases, except near the crust-core
transition, cf. \Fig{fig:miniSly4}). Note that $r_p^{\eff}$ coincides
with $r_p$ in the case $d=1$, but not in the cases $d=2$ or $3$. If we
used an asymmetric surface (see discussion below
\Eq{eq:density_profile}), $r_p^{\eff}$ would differ from $r_p$ also in
the case $d=1$.

Since we define the reference cluster with a common surface at
$r_p^{\eff}$ for protons and neutrons, i.e., without neutron skin, the
reference system contains less neutrons than the actual WS
cell. Therefore, rather than comparing the energies of two systems
having different numbers of particles, one should compare their
thermodynamic potentials. The surface contribution to the
thermodynamic potential, $\Omega_s$, is defined as the change in
$\Omega$ (excluding Coulomb) with respect to the ``bulk''
thermodynamic potential (i.e., excluding gradient terms) of the
reference cluster. We denote $V^{\liq} = V_d(r_p^{\eff})$ the volume
of the reference cluster and $V^{\gas} = V_{\WS}-V^{\liq}$ the volume
of the surrounding gas. Then the surface potential can be written as
\begin{align}
  \Omega_s =& E_{\Sk} - \mu_n N - \mu_p Z \nonumber \\
    &- V^{\liq} (\epsilon_{\Sk}^{\liq} - \mu_n \rho_n^{\liq}
      - \mu_p \rho_p^{\liq}) \nonumber \\
    &-V^{\gas} (\epsilon_{\Sk}^{\gas} - \mu_n \rho_n^{\gas}
      - \mu_p \rho_p^{\gas})\,,
  \label{eq:omega_s}
\end{align}
where $\epsilon_{\Sk}^i$ is the energy density obtained with the
Skyrme functional $\cal{H}$ in the case of uniform matter with
densities $\rho_n^i$ and $\rho_p^i$.

Analogously to \Eq{eq:defrpeff}, we define an effective neutron radius
$r_n^{\eff}$, and an effective volume of the neutron skin $V_s =
V_d(r_n^{\eff})-V_d(r_p^{\eff})$. Then the number of neutrons in the
skin, $N_s$, is given by
\begin{equation}
  N_s = V_s (\rho_n^{\liq}-\rho_n^{\gas})\,,
\end{equation}
and \Eq{eq:omega_s} can be rewritten as
\begin{equation}
  \Omega_s = E_{\Sk} - V^{\liq} \epsilon_{\Sk}^{\liq} - V^{\gas}\epsilon_{\Sk}^{\gas}
    - \mu_n N_s\,.
\end{equation}
Finally, the surface energy $E_s$ is given by $E_s = \Omega_s + \mu_n
N_s$ \cite{Douchin2000}.

\begin{figure}
  \includegraphics[width=8cm]{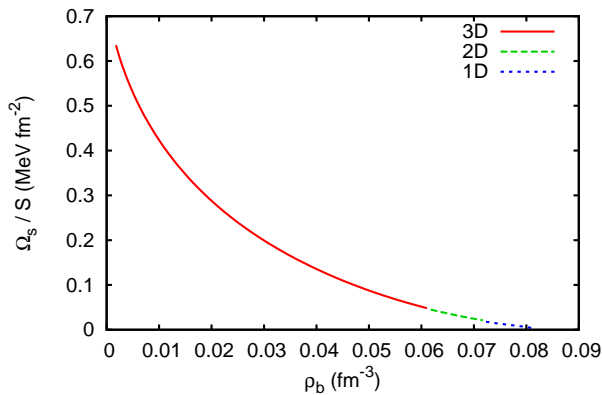}
  \caption{(Color online) The surface tension $\sigma$ as a function of the baryon
    density $\rho_b$. For each $\rho_b$, the result corresponding to
    the most favored geometry is displayed.}
  \label{fig:surf_en}
\end{figure}

The surface tension is approximately given by $\sigma =
\Omega_s/S_d(r_p^{\eff})$, where $S_d(r)$ is defined in \Eq{eq:s_d}.
Although $V_d(r)$ and $S_d(r)$ have the dimensions of a volume and of
an area only in the case $d=3$, the ratio $\Omega_s/S_d(r_p^{\eff})$
is always an energy per area. Note that the above definition of the
surface tension is only exact in 1D or if the cluster radius is so
large that curvature effects can be neglected. In
Fig.~\ref{fig:surf_en}
%%%%%%%%%%%%%%%%%%%%%%%%%%%%%%%%%%%%%%%%%%%%%%%%%%%%%%%%%%%%%%%%%%%%%%%%%%%%%%%
%%%%%%%%%%%%%%%%%%%%%%%%%%%%%%%%%%%%%%%%%%%%%%%%%%%%%%%%%%%%%%%%%%%%%%%%%%%%%%%
we display the surface tension. We see that it decreases with increasing
$\rho_b$. This is not surprising because with increasing $\rho_b$ the densities
in the gas and in the liquid get closer to each other. In the low-density limit,
where we have essentially isolated nuclei, our surface tension is consistent
with the surface energy in the Bethe-Weizs{\"a}cker semi-empirical mass formula
from which one obtains $\sigma = 1.03$ MeV fm$^{-2}$ \cite{RingSchuck}. Our
surface tension is similar to the RMF results shown in Fig. 6(c) of
\cite{Avancini2008}.

\begin{figure}
  \includegraphics[width=8cm]{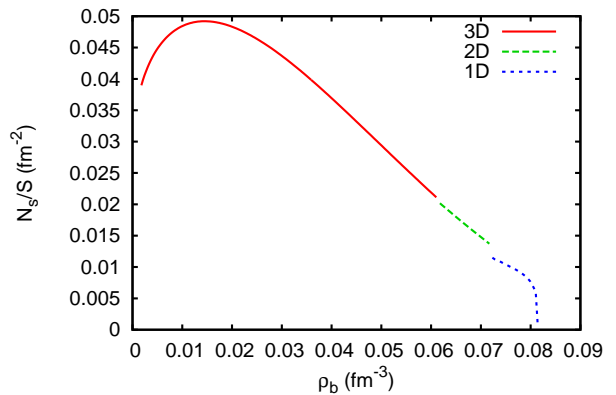}
  \caption{(Color online) The number of the additional neutrons due to the skin,
    $N_s$, divided by the cluster surface $S_d(r_p^{\eff})$.}
  \label{fig:neutron_skin}
\end{figure}

Another interesting quantity is the number of additional neutrons in
the skin. Again, we normalize it to the surface and display in
Fig.~\ref{fig:neutron_skin}
%%%%%%%%%%%%%%%%%%%%%%%%%%%%%%%%%%%%%%%%%%%%%%%%%%%%%%%%%%%%%%%%%%%%%%%%%%%%%%%
%%%%%%%%%%%%%%%%%%%%%%%%%%%%%%%%%%%%%%%%%%%%%%%%%%%%%%%%%%%%%%%%%%%%%%%%%%%%%%%
the number of skin neutrons per unit surface,
$N_s/S_d(r_p^{\eff})$. We observe that, after a maximum at $\rho_b
\sim 0.014~\text{fm}^{-3}$, the number of neutrons in the skin
decreases with increasing baryon density, although the skin thickness
remains about $s_n \sim 0.45~\text{fm}$, because the density
difference $\rho_n^{\liq}-\rho_n^{\gas}$ decreases. Note that, compared
to the total number of neutrons in the WS cell, the number of neutrons
in the skin is very small: for instance, for $\rho_b =
0.02~\text{fm}^{-3}$ there are about $800$ neutrons in the cell, $200$
in the cluster, and only $23$ in the skin.

%%%%%%%%%%%%%%%%%%%%%%%%%%%%%%%%%%%%%%%%%%%%%%%%%%%%%%%%%%%%%%%%%%%%%%%%%%%%%%%
\section{Conclusions}
\label{sec:conclusions}
%%%%%%%%%%%%%%%%%%%%%%%%%%%%%%%%%%%%%%%%%%%%%%%%%%%%%%%%%%%%%%%%%%%%%%%%%%%%%%%
In our work we compared the results obtained for the inner crust
within a simple phase-coexistence picture with results of the
minimization of the energy with respect to the density profile. We
used Skyrme interactions and the ETF approximation to calculate the
energy. Both approaches give similar results for the neutron and
proton densities in the clusters and in the gas, despite the Coulomb
and surface effects included in the minimization but not in the phase
coexistence.

However the phase-coexistence picture is insufficient to predict the
cluster sizes and transitions between different geometries, since
these result from the competition of Coulomb and surface energies not
included in the phase coexistence. With increasing baryon density, we
find crystals, rods and plates, but no ``inverted'' geometries such as
tubes and bubbles. Because of the small proton fraction in $\beta$
equilibrium, the energy differences between the different geometries
are extremely small. Another effect of Coulomb and surface energies is
to shift the crust-core transition to lower baryon density.

Although the densities obtained by energy minimization do not present
strong deviations from those of the phase coexistence, the proton
fraction does, especially at low baryon densities. The small
corrections to mechanical and chemical equilibrium due to Coulomb and
surface effects slightly modify the density of the neutron gas,
resulting in a considerable reduction of volume and proton fractions.
The proton fractions obtained within the minimization are similar to
HFB results in the literature \cite{Pearson2014}, although shell
effects, which are present in HFB, are missing in the ETF
approximation to the energy.

The minimization allowed us to calculate the surface tension and the
number of skin neutrons in the cluster surface. We plan to use these
results in an improved hydrodynamic description of collective modes in
the inner crust similar to the one of \Ref{DiGallo2011} but including
Coulomb and surface effects.

% REFERENCES
  \nocite*
  \bibliographystyle{apsrev4-1}
  \bibliography{coex}

\begin{thebibliography}{35}%
\makeatletter
\providecommand \@ifxundefined [1]{%
 \@ifx{#1\undefined}
}%
\providecommand \@ifnum [1]{%
 \ifnum #1\expandafter \@firstoftwo
 \else \expandafter \@secondoftwo
 \fi
}%
\providecommand \@ifx [1]{%
 \ifx #1\expandafter \@firstoftwo
 \else \expandafter \@secondoftwo
 \fi
}%
\providecommand \natexlab [1]{#1}%
\providecommand \enquote  [1]{``#1''}%
\providecommand \bibnamefont  [1]{#1}%
\providecommand \bibfnamefont [1]{#1}%
\providecommand \citenamefont [1]{#1}%
\providecommand \href@noop [0]{\@secondoftwo}%
\providecommand \href [0]{\begingroup \@sanitize@url \@href}%
\providecommand \@href[1]{\@@startlink{#1}\@@href}%
\providecommand \@@href[1]{\endgroup#1\@@endlink}%
\providecommand \@sanitize@url [0]{\catcode `\\12\catcode `\$12\catcode
  `\&12\catcode `\#12\catcode `\^12\catcode `\_12\catcode `\%12\relax}%
\providecommand \@@startlink[1]{}%
\providecommand \@@endlink[0]{}%
\providecommand \url  [0]{\begingroup\@sanitize@url \@url }%
\providecommand \@url [1]{\endgroup\@href {#1}{\urlprefix }}%
\providecommand \urlprefix  [0]{URL }%
\providecommand \Eprint [0]{\href }%
\providecommand \doibase [0]{http://dx.doi.org/}%
\providecommand \selectlanguage [0]{\@gobble}%
\providecommand \bibinfo  [0]{\@secondoftwo}%
\providecommand \bibfield  [0]{\@secondoftwo}%
\providecommand \translation [1]{[#1]}%
\providecommand \BibitemOpen [0]{}%
\providecommand \bibitemStop [0]{}%
\providecommand \bibitemNoStop [0]{.\EOS\space}%
\providecommand \EOS [0]{\spacefactor3000\relax}%
\providecommand \BibitemShut  [1]{\csname bibitem#1\endcsname}%
\let\auto@bib@innerbib\@empty
%</preamble>
\bibitem [{\citenamefont {Chamel}\ and\ \citenamefont
  {Haensel}(2008)}]{Chamel2008}%
  \BibitemOpen
  \bibfield  {author} {\bibinfo {author} {\bibfnamefont {N.}~\bibnamefont
  {Chamel}}\ and\ \bibinfo {author} {\bibfnamefont {P.}~\bibnamefont
  {Haensel}},\ }\href@noop {} {\bibfield  {journal} {\bibinfo  {journal}
  {Living Reviews in Relativity}\ }\textbf {\bibinfo {volume} {11}},\ \bibinfo
  {pages} {10} (\bibinfo {year} {2008})}\BibitemShut {NoStop}%
\bibitem [{\citenamefont {Ravenhall}\ \emph {et~al.}(1983)\citenamefont
  {Ravenhall}, \citenamefont {Pethick},\ and\ \citenamefont
  {Wilson}}]{Ravenhall1983}%
  \BibitemOpen
  \bibfield  {author} {\bibinfo {author} {\bibfnamefont {D.~G.}\ \bibnamefont
  {Ravenhall}}, \bibinfo {author} {\bibfnamefont {C.~J.}\ \bibnamefont
  {Pethick}}, \ and\ \bibinfo {author} {\bibfnamefont {J.~R.}\ \bibnamefont
  {Wilson}},\ }\href {\doibase 10.1103/PhysRevLett.50.2066} {\bibfield
  {journal} {\bibinfo  {journal} {Phys. Rev. Lett.}\ }\textbf {\bibinfo
  {volume} {50}},\ \bibinfo {pages} {2066} (\bibinfo {year}
  {1983})}\BibitemShut {NoStop}%
\bibitem [{\citenamefont {Hashimoto}\ \emph {et~al.}(1984)\citenamefont
  {Hashimoto}, \citenamefont {Seki},\ and\ \citenamefont
  {Yamada}}]{Hashimoto1984}%
  \BibitemOpen
  \bibfield  {author} {\bibinfo {author} {\bibfnamefont {M.-a.}\ \bibnamefont
  {Hashimoto}}, \bibinfo {author} {\bibfnamefont {H.}~\bibnamefont {Seki}}, \
  and\ \bibinfo {author} {\bibfnamefont {M.}~\bibnamefont {Yamada}},\ }\href
  {\doibase 10.1143/PTP.71.320} {\bibfield  {journal} {\bibinfo  {journal}
  {Prog. Theor. Phys.}\ }\textbf {\bibinfo {volume} {71}},\ \bibinfo {pages}
  {320} (\bibinfo {year} {1984})}\BibitemShut {NoStop}%
\bibitem [{\citenamefont {Lassaut}\ \emph {et~al.}(1987)\citenamefont
  {Lassaut}, \citenamefont {Flocard}, \citenamefont {Bonche}, \citenamefont
  {Heenen},\ and\ \citenamefont {Suraud}}]{Lassaut1987}%
  \BibitemOpen
  \bibfield  {author} {\bibinfo {author} {\bibfnamefont {M.}~\bibnamefont
  {Lassaut}}, \bibinfo {author} {\bibfnamefont {H.}~\bibnamefont {Flocard}},
  \bibinfo {author} {\bibfnamefont {P.}~\bibnamefont {Bonche}}, \bibinfo
  {author} {\bibfnamefont {P.}~\bibnamefont {Heenen}}, \ and\ \bibinfo {author}
  {\bibfnamefont {E.}~\bibnamefont {Suraud}},\ }\href@noop {} {\bibfield
  {journal} {\bibinfo  {journal} {Astron. Astrophys.}\ }\textbf {\bibinfo
  {volume} {198}},\ \bibinfo {pages} {L3} (\bibinfo {year} {1987})}\BibitemShut
  {NoStop}%
\bibitem [{\citenamefont {Oyamatsu}(1993)}]{Oyamatsu1993}%
  \BibitemOpen
  \bibfield  {author} {\bibinfo {author} {\bibfnamefont {K.}~\bibnamefont
  {Oyamatsu}},\ }\href@noop {} {\bibfield  {journal} {\bibinfo  {journal}
  {Nucl. Phys. A}\ }\textbf {\bibinfo {volume} {561}},\ \bibinfo {pages} {431 }
  (\bibinfo {year} {1993})}\BibitemShut {NoStop}%
\bibitem [{\citenamefont {Watanabe}\ \emph {et~al.}(2003)\citenamefont
  {Watanabe}, \citenamefont {Sato}, \citenamefont {Yasuoka},\ and\
  \citenamefont {Ebisuzaki}}]{Watanabe2003}%
  \BibitemOpen
  \bibfield  {author} {\bibinfo {author} {\bibfnamefont {G.}~\bibnamefont
  {Watanabe}}, \bibinfo {author} {\bibfnamefont {K.}~\bibnamefont {Sato}},
  \bibinfo {author} {\bibfnamefont {K.}~\bibnamefont {Yasuoka}}, \ and\
  \bibinfo {author} {\bibfnamefont {T.}~\bibnamefont {Ebisuzaki}},\ }\href@noop
  {} {\bibfield  {journal} {\bibinfo  {journal} {Phys. Rev. C}\ }\textbf
  {\bibinfo {volume} {68}},\ \bibinfo {pages} {035806} (\bibinfo {year}
  {2003})}\BibitemShut {NoStop}%
\bibitem [{\citenamefont {Napolitani}\ \emph {et~al.}(2007)\citenamefont
  {Napolitani}, \citenamefont {Chomaz}, \citenamefont {Gulminelli},\ and\
  \citenamefont {Hasnaoui}}]{Napolitani2007}%
  \BibitemOpen
  \bibfield  {author} {\bibinfo {author} {\bibfnamefont {P.}~\bibnamefont
  {Napolitani}}, \bibinfo {author} {\bibfnamefont {P.}~\bibnamefont {Chomaz}},
  \bibinfo {author} {\bibfnamefont {F.}~\bibnamefont {Gulminelli}}, \ and\
  \bibinfo {author} {\bibfnamefont {K.~H.~O.}\ \bibnamefont {Hasnaoui}},\
  }\href@noop {} {\bibfield  {journal} {\bibinfo  {journal} {Phys. Rev. Lett.}\
  }\textbf {\bibinfo {volume} {98}},\ \bibinfo {pages} {131102} (\bibinfo
  {year} {2007})}\BibitemShut {NoStop}%
\bibitem [{\citenamefont {Avancini}\ \emph {et~al.}(2008)\citenamefont
  {Avancini}, \citenamefont {Menezes}, \citenamefont {Alloy}, \citenamefont
  {Marinelli}, \citenamefont {Moraes},\ and\ \citenamefont
  {Provid\^encia}}]{Avancini2008}%
  \BibitemOpen
  \bibfield  {author} {\bibinfo {author} {\bibfnamefont {S.~S.}\ \bibnamefont
  {Avancini}}, \bibinfo {author} {\bibfnamefont {D.~P.}\ \bibnamefont
  {Menezes}}, \bibinfo {author} {\bibfnamefont {M.~D.}\ \bibnamefont {Alloy}},
  \bibinfo {author} {\bibfnamefont {J.~R.}\ \bibnamefont {Marinelli}}, \bibinfo
  {author} {\bibfnamefont {M.~M.~W.}\ \bibnamefont {Moraes}}, \ and\ \bibinfo
  {author} {\bibfnamefont {C.}~\bibnamefont {Provid\^encia}},\ }\href {\doibase
  10.1103/PhysRevC.78.015802} {\bibfield  {journal} {\bibinfo  {journal} {Phys.
  Rev. C}\ }\textbf {\bibinfo {volume} {78}},\ \bibinfo {pages} {015802}
  (\bibinfo {year} {2008})}\BibitemShut {NoStop}%
\bibitem [{\citenamefont {Avancini}\ \emph {et~al.}(2010)\citenamefont
  {Avancini}, \citenamefont {Chiacchiera}, \citenamefont {Menezes},\ and\
  \citenamefont {Provid\^encia}}]{Avancini2010}%
  \BibitemOpen
  \bibfield  {author} {\bibinfo {author} {\bibfnamefont {S.~S.}\ \bibnamefont
  {Avancini}}, \bibinfo {author} {\bibfnamefont {S.}~\bibnamefont
  {Chiacchiera}}, \bibinfo {author} {\bibfnamefont {D.~P.}\ \bibnamefont
  {Menezes}}, \ and\ \bibinfo {author} {\bibfnamefont {C.}~\bibnamefont
  {Provid\^encia}},\ }\href {\doibase 10.1103/PhysRevC.82.055807} {\bibfield
  {journal} {\bibinfo  {journal} {Phys. Rev. C}\ }\textbf {\bibinfo {volume}
  {82}},\ \bibinfo {pages} {055807} (\bibinfo {year} {2010})}\BibitemShut
  {NoStop}%
\bibitem [{\citenamefont {Sebille}\ \emph {et~al.}(2011)\citenamefont
  {Sebille}, \citenamefont {de~la Mota},\ and\ \citenamefont
  {Figerou}}]{Sebille2011}%
  \BibitemOpen
  \bibfield  {author} {\bibinfo {author} {\bibfnamefont {F.}~\bibnamefont
  {Sebille}}, \bibinfo {author} {\bibfnamefont {V.}~\bibnamefont {de~la Mota}},
  \ and\ \bibinfo {author} {\bibfnamefont {S.}~\bibnamefont {Figerou}},\
  }\href@noop {} {\bibfield  {journal} {\bibinfo  {journal} {Phys. Rev. C}\
  }\textbf {\bibinfo {volume} {84}},\ \bibinfo {pages} {055801} (\bibinfo
  {year} {2011})}\BibitemShut {NoStop}%
\bibitem [{\citenamefont {Baldo}\ \emph {et~al.}(2014)\citenamefont {Baldo},
  \citenamefont {Burgio}, \citenamefont {Centelles}, \citenamefont {Sharma},\
  and\ \citenamefont {Vi\~nas}}]{Baldo2014}%
  \BibitemOpen
  \bibfield  {author} {\bibinfo {author} {\bibfnamefont {M.}~\bibnamefont
  {Baldo}}, \bibinfo {author} {\bibfnamefont {G.}~\bibnamefont {Burgio}},
  \bibinfo {author} {\bibfnamefont {M.}~\bibnamefont {Centelles}}, \bibinfo
  {author} {\bibfnamefont {B.}~\bibnamefont {Sharma}}, \ and\ \bibinfo {author}
  {\bibfnamefont {X.}~\bibnamefont {Vi\~nas}},\ }\href@noop {} {\bibfield
  {journal} {\bibinfo  {journal} {Yad. Fiz. [Sov. J. Nucl. Phys.]}\ }\textbf
  {\bibinfo {volume} {77}},\ \bibinfo {pages} {1157} (\bibinfo {year}
  {2014})}\BibitemShut {NoStop}%
\bibitem [{\citenamefont {Pais}\ \emph {et~al.}(2014)\citenamefont {Pais},
  \citenamefont {Newton},\ and\ \citenamefont {Stone}}]{Pais2014}%
  \BibitemOpen
  \bibfield  {author} {\bibinfo {author} {\bibfnamefont {H.}~\bibnamefont
  {Pais}}, \bibinfo {author} {\bibfnamefont {W.~G.}\ \bibnamefont {Newton}}, \
  and\ \bibinfo {author} {\bibfnamefont {J.~R.}\ \bibnamefont {Stone}},\ }\href
  {\doibase 10.1103/PhysRevC.90.065802} {\bibfield  {journal} {\bibinfo
  {journal} {Phys. Rev. C}\ }\textbf {\bibinfo {volume} {90}},\ \bibinfo
  {pages} {065802} (\bibinfo {year} {2014})}\BibitemShut {NoStop}%
\bibitem [{\citenamefont {Pais}\ \emph {et~al.}(2015)\citenamefont {Pais},
  \citenamefont {Chiacchiera},\ and\ \citenamefont {Provid\^encia}}]{Pais2015}%
  \BibitemOpen
  \bibfield  {author} {\bibinfo {author} {\bibfnamefont {H.}~\bibnamefont
  {Pais}}, \bibinfo {author} {\bibfnamefont {S.}~\bibnamefont {Chiacchiera}}, \
  and\ \bibinfo {author} {\bibfnamefont {C.}~\bibnamefont {Provid\^encia}},\
  }\href@noop {} {\bibfield  {journal} {\bibinfo  {journal} {arXiv e-print}\ }
  (\bibinfo {year} {2015})},\ \Eprint {http://arxiv.org/abs/1504.03964}
  {arXiv:1504.03964 [nucl-th]} \BibitemShut {NoStop}%
\bibitem [{\citenamefont {Aymard}\ \emph {et~al.}(2014)\citenamefont {Aymard},
  \citenamefont {Gulminelli},\ and\ \citenamefont {Margueron}}]{Aymard2014}%
  \BibitemOpen
  \bibfield  {author} {\bibinfo {author} {\bibfnamefont {F.}~\bibnamefont
  {Aymard}}, \bibinfo {author} {\bibfnamefont {F.}~\bibnamefont {Gulminelli}},
  \ and\ \bibinfo {author} {\bibfnamefont {J.}~\bibnamefont {Margueron}},\
  }\href@noop {} {\bibfield  {journal} {\bibinfo  {journal} {Phys. Rev. C}\
  }\textbf {\bibinfo {volume} {89}},\ \bibinfo {pages} {065807} (\bibinfo
  {year} {2014})}\BibitemShut {NoStop}%
\bibitem [{\citenamefont {Magierski}(2004)}]{Magierski2004}%
  \BibitemOpen
  \bibfield  {author} {\bibinfo {author} {\bibfnamefont {P.}~\bibnamefont
  {Magierski}},\ }\href@noop {} {\bibfield  {journal} {\bibinfo  {journal}
  {Int. J. Mod. Phys. E}\ }\textbf {\bibinfo {volume} {13}},\ \bibinfo {pages}
  {371} (\bibinfo {year} {2004})}\BibitemShut {NoStop}%
\bibitem [{\citenamefont {Magierski}\ and\ \citenamefont
  {Bulgac}(2004)}]{MagierskiBulgac2004}%
  \BibitemOpen
  \bibfield  {author} {\bibinfo {author} {\bibfnamefont {P.}~\bibnamefont
  {Magierski}}\ and\ \bibinfo {author} {\bibfnamefont {A.}~\bibnamefont
  {Bulgac}},\ }\href@noop {} {\bibfield  {journal} {\bibinfo  {journal} {Acta
  Phys. Polon. B}\ }\textbf {\bibinfo {volume} {35}},\ \bibinfo {pages} {1203}
  (\bibinfo {year} {2004})}\BibitemShut {NoStop}%
\bibitem [{\citenamefont {Di~Gallo}\ \emph {et~al.}(2011)\citenamefont
  {Di~Gallo}, \citenamefont {Oertel},\ and\ \citenamefont
  {Urban}}]{DiGallo2011}%
  \BibitemOpen
  \bibfield  {author} {\bibinfo {author} {\bibfnamefont {L.}~\bibnamefont
  {Di~Gallo}}, \bibinfo {author} {\bibfnamefont {M.}~\bibnamefont {Oertel}}, \
  and\ \bibinfo {author} {\bibfnamefont {M.}~\bibnamefont {Urban}},\
  }\href@noop {} {\bibfield  {journal} {\bibinfo  {journal} {Phys. Rev. C}\
  }\textbf {\bibinfo {volume} {84}},\ \bibinfo {pages} {045801} (\bibinfo
  {year} {2011})}\BibitemShut {NoStop}%
\bibitem [{\citenamefont {Page}\ and\ \citenamefont
  {Reddy}(2012)}]{PageReddy2012}%
  \BibitemOpen
  \bibfield  {author} {\bibinfo {author} {\bibfnamefont {D.}~\bibnamefont
  {Page}}\ and\ \bibinfo {author} {\bibfnamefont {S.}~\bibnamefont {Reddy}},\
  }\enquote {\bibinfo {title} {Neutron star crust},}\ \ (\bibinfo  {publisher}
  {Nova Science Publishers},\ \bibinfo {address} {Hauppauge},\ \bibinfo {year}
  {2012})\ Chap.~\bibinfo {chapter} {14}\BibitemShut {NoStop}%
\bibitem [{\citenamefont {Khan}\ \emph {et~al.}(2005)\citenamefont {Khan},
  \citenamefont {Sandulescu},\ and\ \citenamefont {Giai}}]{Khan2005}%
  \BibitemOpen
  \bibfield  {author} {\bibinfo {author} {\bibfnamefont {E.}~\bibnamefont
  {Khan}}, \bibinfo {author} {\bibfnamefont {N.}~\bibnamefont {Sandulescu}}, \
  and\ \bibinfo {author} {\bibfnamefont {N.~V.}\ \bibnamefont {Giai}},\
  }\href@noop {} {\bibfield  {journal} {\bibinfo  {journal} {Phys. Rev. C}\
  }\textbf {\bibinfo {volume} {71}},\ \bibinfo {pages} {042801} (\bibinfo
  {year} {2005})}\BibitemShut {NoStop}%
\bibitem [{\citenamefont {Cirigliano}\ \emph {et~al.}(2011)\citenamefont
  {Cirigliano}, \citenamefont {Reddy},\ and\ \citenamefont
  {Sharma}}]{Cirigliano2011}%
  \BibitemOpen
  \bibfield  {author} {\bibinfo {author} {\bibfnamefont {V.}~\bibnamefont
  {Cirigliano}}, \bibinfo {author} {\bibfnamefont {S.}~\bibnamefont {Reddy}}, \
  and\ \bibinfo {author} {\bibfnamefont {R.}~\bibnamefont {Sharma}},\ }\href
  {\doibase 10.1103/PhysRevC.84.045809} {\bibfield  {journal} {\bibinfo
  {journal} {Phys. Rev. C}\ }\textbf {\bibinfo {volume} {84}},\ \bibinfo
  {pages} {045809} (\bibinfo {year} {2011})}\BibitemShut {NoStop}%
\bibitem [{\citenamefont {Chamel}\ \emph {et~al.}(2013)\citenamefont {Chamel},
  \citenamefont {Page},\ and\ \citenamefont {Reddy}}]{Chamel2013CM}%
  \BibitemOpen
  \bibfield  {author} {\bibinfo {author} {\bibfnamefont {N.}~\bibnamefont
  {Chamel}}, \bibinfo {author} {\bibfnamefont {D.}~\bibnamefont {Page}}, \ and\
  \bibinfo {author} {\bibfnamefont {S.}~\bibnamefont {Reddy}},\ }\href
  {\doibase 10.1103/PhysRevC.87.035803} {\bibfield  {journal} {\bibinfo
  {journal} {Phys. Rev. C}\ }\textbf {\bibinfo {volume} {87}},\ \bibinfo
  {pages} {035803} (\bibinfo {year} {2013})}\BibitemShut {NoStop}%
\bibitem [{\citenamefont {Chabanat}\ \emph {et~al.}(1997)\citenamefont
  {Chabanat}, \citenamefont {Bonche}, \citenamefont {Haensel}, \citenamefont
  {Meyer},\ and\ \citenamefont {Schaeffer}}]{Chabanat1997}%
  \BibitemOpen
  \bibfield  {author} {\bibinfo {author} {\bibfnamefont {E.}~\bibnamefont
  {Chabanat}}, \bibinfo {author} {\bibfnamefont {P.}~\bibnamefont {Bonche}},
  \bibinfo {author} {\bibfnamefont {P.}~\bibnamefont {Haensel}}, \bibinfo
  {author} {\bibfnamefont {J.}~\bibnamefont {Meyer}}, \ and\ \bibinfo {author}
  {\bibfnamefont {R.}~\bibnamefont {Schaeffer}},\ }\href@noop {} {\bibfield
  {journal} {\bibinfo  {journal} {Nucl. Phys. A}\ }\textbf {\bibinfo {volume}
  {627}},\ \bibinfo {pages} {710 } (\bibinfo {year} {1997})}\BibitemShut
  {NoStop}%
\bibitem [{\citenamefont {Chabanat}\ \emph {et~al.}(1998)\citenamefont
  {Chabanat}, \citenamefont {Bonche}, \citenamefont {Haensel}, \citenamefont
  {Meyer},\ and\ \citenamefont {Schaeffer}}]{Chabanat1998}%
  \BibitemOpen
  \bibfield  {author} {\bibinfo {author} {\bibfnamefont {E.}~\bibnamefont
  {Chabanat}}, \bibinfo {author} {\bibfnamefont {P.}~\bibnamefont {Bonche}},
  \bibinfo {author} {\bibfnamefont {P.}~\bibnamefont {Haensel}}, \bibinfo
  {author} {\bibfnamefont {J.}~\bibnamefont {Meyer}}, \ and\ \bibinfo {author}
  {\bibfnamefont {R.}~\bibnamefont {Schaeffer}},\ }\href@noop {} {\bibfield
  {journal} {\bibinfo  {journal} {Nucl. Phys. A}\ }\textbf {\bibinfo {volume}
  {635}},\ \bibinfo {pages} {231 } (\bibinfo {year} {1998})}\BibitemShut
  {NoStop}%
\bibitem [{\citenamefont {Goriely}\ \emph {et~al.}(2013)\citenamefont
  {Goriely}, \citenamefont {Chamel},\ and\ \citenamefont
  {Pearson}}]{Chamel2013}%
  \BibitemOpen
  \bibfield  {author} {\bibinfo {author} {\bibfnamefont {S.}~\bibnamefont
  {Goriely}}, \bibinfo {author} {\bibfnamefont {N.}~\bibnamefont {Chamel}}, \
  and\ \bibinfo {author} {\bibfnamefont {J.~M.}\ \bibnamefont {Pearson}},\
  }\href {\doibase 10.1103/PhysRevC.88.024308} {\bibfield  {journal} {\bibinfo
  {journal} {Phys. Rev. C}\ }\textbf {\bibinfo {volume} {88}},\ \bibinfo
  {pages} {024308} (\bibinfo {year} {2013})}\BibitemShut {NoStop}%
\bibitem [{\citenamefont {Brack}\ \emph {et~al.}(1985)\citenamefont {Brack},
  \citenamefont {Guet},\ and\ \citenamefont {H\r{a}kansson}}]{Brack1985}%
  \BibitemOpen
  \bibfield  {author} {\bibinfo {author} {\bibfnamefont {M.}~\bibnamefont
  {Brack}}, \bibinfo {author} {\bibfnamefont {C.}~\bibnamefont {Guet}}, \ and\
  \bibinfo {author} {\bibfnamefont {H.-B.}\ \bibnamefont {H\r{a}kansson}},\
  }\href {\doibase http://dx.doi.org/10.1016/0370-1573(86)90078-5} {\bibfield
  {journal} {\bibinfo  {journal} {Phys. Rep.}\ }\textbf {\bibinfo {volume}
  {123}},\ \bibinfo {pages} {275 } (\bibinfo {year} {1985})}\BibitemShut
  {NoStop}%
\bibitem [{\citenamefont {Negele}\ and\ \citenamefont
  {Vautherin}(1973)}]{Negele1973}%
  \BibitemOpen
  \bibfield  {author} {\bibinfo {author} {\bibfnamefont {J.}~\bibnamefont
  {Negele}}\ and\ \bibinfo {author} {\bibfnamefont {D.}~\bibnamefont
  {Vautherin}},\ }\href@noop {} {\bibfield  {journal} {\bibinfo  {journal}
  {Nucl. Phys. A}\ }\textbf {\bibinfo {volume} {207}},\ \bibinfo {pages} {298 }
  (\bibinfo {year} {1973})}\BibitemShut {NoStop}%
\bibitem [{\citenamefont {Pearson}\ \emph {et~al.}(2012)\citenamefont
  {Pearson}, \citenamefont {Chamel}, \citenamefont {Goriely},\ and\
  \citenamefont {Ducoin}}]{Pearson2012}%
  \BibitemOpen
  \bibfield  {author} {\bibinfo {author} {\bibfnamefont {J.~M.}\ \bibnamefont
  {Pearson}}, \bibinfo {author} {\bibfnamefont {N.}~\bibnamefont {Chamel}},
  \bibinfo {author} {\bibfnamefont {S.}~\bibnamefont {Goriely}}, \ and\
  \bibinfo {author} {\bibfnamefont {C.}~\bibnamefont {Ducoin}},\ }\href
  {\doibase 10.1103/PhysRevC.85.065803} {\bibfield  {journal} {\bibinfo
  {journal} {Phys. Rev. C}\ }\textbf {\bibinfo {volume} {85}},\ \bibinfo
  {pages} {065803} (\bibinfo {year} {2012})}\BibitemShut {NoStop}%
\bibitem [{\citenamefont {Maruyama}\ \emph {et~al.}(2005)\citenamefont
  {Maruyama}, \citenamefont {Tatsumi}, \citenamefont {Voskresensky},
  \citenamefont {Tanigawa},\ and\ \citenamefont {Chiba}}]{Maruyama2005}%
  \BibitemOpen
  \bibfield  {author} {\bibinfo {author} {\bibfnamefont {T.}~\bibnamefont
  {Maruyama}}, \bibinfo {author} {\bibfnamefont {T.}~\bibnamefont {Tatsumi}},
  \bibinfo {author} {\bibfnamefont {D.~N.}\ \bibnamefont {Voskresensky}},
  \bibinfo {author} {\bibfnamefont {T.}~\bibnamefont {Tanigawa}}, \ and\
  \bibinfo {author} {\bibfnamefont {S.}~\bibnamefont {Chiba}},\ }\href
  {\doibase 10.1103/PhysRevC.72.015802} {\bibfield  {journal} {\bibinfo
  {journal} {Phys. Rev. C}\ }\textbf {\bibinfo {volume} {72}},\ \bibinfo
  {pages} {015802} (\bibinfo {year} {2005})}\BibitemShut {NoStop}%
\bibitem [{\citenamefont {Salpeter}(1961)}]{Salpeter1961}%
  \BibitemOpen
  \bibfield  {author} {\bibinfo {author} {\bibfnamefont {E.}~\bibnamefont
  {Salpeter}},\ }\href@noop {} {\bibfield  {journal} {\bibinfo  {journal}
  {Astrophys. J.}\ ,\ \bibinfo {pages} {669}} (\bibinfo {year}
  {1961})}\BibitemShut {NoStop}%
\bibitem [{\citenamefont {Rajagopal}(1978)}]{Rajagopal1978}%
  \BibitemOpen
  \bibfield  {author} {\bibinfo {author} {\bibfnamefont {A.~K.}\ \bibnamefont
  {Rajagopal}},\ }\href@noop {} {\bibfield  {journal} {\bibinfo  {journal} {J.
  Phys. C}\ }\textbf {\bibinfo {volume} {11}} (\bibinfo {year}
  {1978})}\BibitemShut {NoStop}%
\bibitem [{\citenamefont {Piekarewicz}\ and\ \citenamefont
  {S\'anchez}(2012)}]{Piekarewicz2012}%
  \BibitemOpen
  \bibfield  {author} {\bibinfo {author} {\bibfnamefont {J.}~\bibnamefont
  {Piekarewicz}}\ and\ \bibinfo {author} {\bibfnamefont {G.~T.}\ \bibnamefont
  {S\'anchez}},\ }\href {\doibase 10.1103/PhysRevC.85.015807} {\bibfield
  {journal} {\bibinfo  {journal} {Phys. Rev. C}\ }\textbf {\bibinfo {volume}
  {85}},\ \bibinfo {pages} {015807} (\bibinfo {year} {2012})}\BibitemShut
  {NoStop}%
\bibitem [{\citenamefont {Lattimer}\ and\ \citenamefont
  {Swesty}(1991)}]{Lattimer1991}%
  \BibitemOpen
  \bibfield  {author} {\bibinfo {author} {\bibfnamefont {J.~M.}\ \bibnamefont
  {Lattimer}}\ and\ \bibinfo {author} {\bibfnamefont {F.~D.}\ \bibnamefont
  {Swesty}},\ }\href {\doibase http://dx.doi.org/10.1016/0375-9474(91)90452-C}
  {\bibfield  {journal} {\bibinfo  {journal} {Nucl. Phys. A}\ }\textbf
  {\bibinfo {volume} {535}},\ \bibinfo {pages} {331 } (\bibinfo {year}
  {1991})}\BibitemShut {NoStop}%
\bibitem [{\citenamefont {Pearson}\ \emph {et~al.}(2014)\citenamefont
  {Pearson}, \citenamefont {Chamel}, \citenamefont {Fantina},\ and\
  \citenamefont {Goriely}}]{Pearson2014}%
  \BibitemOpen
  \bibfield  {author} {\bibinfo {author} {\bibfnamefont {J.}~\bibnamefont
  {Pearson}}, \bibinfo {author} {\bibfnamefont {N.}~\bibnamefont {Chamel}},
  \bibinfo {author} {\bibfnamefont {A.}~\bibnamefont {Fantina}}, \ and\
  \bibinfo {author} {\bibfnamefont {S.}~\bibnamefont {Goriely}},\ }\href
  {http://dx.doi.org/10.1140/epja/i2014-14043-8} {\bibfield  {journal}
  {\bibinfo  {journal} {Eur. Phys. J. A}\ }\textbf {\bibinfo {volume} {50}},\
  \bibinfo {eid} {43} (\bibinfo {year} {2014})}\BibitemShut {NoStop}%
\bibitem [{\citenamefont {Douchin}\ \emph {et~al.}(2000)\citenamefont
  {Douchin}, \citenamefont {Haensel},\ and\ \citenamefont
  {Meyer}}]{Douchin2000}%
  \BibitemOpen
  \bibfield  {author} {\bibinfo {author} {\bibfnamefont {F.}~\bibnamefont
  {Douchin}}, \bibinfo {author} {\bibfnamefont {P.}~\bibnamefont {Haensel}}, \
  and\ \bibinfo {author} {\bibfnamefont {J.}~\bibnamefont {Meyer}},\
  }\href@noop {} {\bibfield  {journal} {\bibinfo  {journal} {Nucl. Phys. A}\
  }\textbf {\bibinfo {volume} {665}},\ \bibinfo {pages} {419 } (\bibinfo {year}
  {2000})}\BibitemShut {NoStop}%
\bibitem [{\citenamefont {Ring}\ and\ \citenamefont
  {Schuck}(1980)}]{RingSchuck}%
  \BibitemOpen
  \bibfield  {author} {\bibinfo {author} {\bibfnamefont {P.}~\bibnamefont
  {Ring}}\ and\ \bibinfo {author} {\bibfnamefont {P.}~\bibnamefont {Schuck}},\
  }\href@noop {} {\emph {\bibinfo {title} {The Nuclear Many-Body Problem}}}\
  (\bibinfo  {publisher} {Springer, Berlin},\ \bibinfo {year}
  {1980})\BibitemShut {NoStop}%
\end{thebibliography}%
\end{document}